\documentclass[a4paper,11pt]{article}
\pdfoutput=1 

\usepackage{jcappub} 
\usepackage{bm} 

\usepackage[T1]{fontenc} 

\usepackage[normalem]{ulem}
\usepackage{xcolor}

\usepackage[T1]{fontenc} 

\usepackage{hyperref}

\usepackage{lscape} 

\title{Prospects on the Detection of Solar Dark Photons by the International Axion Observatory}

\author[a,b,1]{T. O'Shea,\note{Corresponding author.}}
\author[a,b,c]{M. Giannotti,}
\author[a,b]{I. G. Irastorza,}
\author[a,b]{L. M. Plasencia,}
\author[a,b]{J. Redondo,}
\author[a,b]{J. Ruz,}
\author[a,b]{and J. K. Vogel}

\affiliation[a]{Departamento de Física Teórica, Universidad de Zaragoza, Zaragoza, 50009, Spain}

\affiliation[b]{Centro de Astropart{\'i}culas y F{\'i}sica de Altas Energ{\'i}as (CAPA), Universidad de Zaragoza, Zaragoza, 50009, Spain}

\affiliation[c]{Department of Chemistry and Physics, Barry University, 11300 NE 2nd Ave., Miami Shores, FL 33161, USA}

\emailAdd{tgerard@unizar.es}
\emailAdd{mgiannotti@unizar.es}
\emailAdd{Igor.Irastorza@cern.ch}
\emailAdd{plasencialuismiguel@gmail.com}
\emailAdd{jredondo@unizar.es}
\emailAdd{jruz@unizar.es}
\emailAdd{jvogel@unizar.es}

\abstract{
Dark (hidden) photons are widely recognised as well motivated candidates for physics beyond the standard model, and have been invoked for the solution  of several outstanding problems, including to account for the dark matter in the universe.
In this paper, we consider a simple model for dark photons, which is coupled to ordinary matter only through kinetic mixing with ordinary photons. 
Within this framework, we calculate the flux of solar dark photons on Earth and revise the potential to detect it with the next generation of  axion helioscopes, particularly with the International AXion Observatory (IAXO). 
This paper extends on previous theoretical analyses in two main ways.
Firstly, it includes a more complete analysis of the possible sources of dark photons from the sun, including the contribution of the solar magnetic field and of nuclear processes,
and secondly 
it includes predictions on the parameter space accessible in the gas-filled phase of IAXO.
}

\begin{document}
\maketitle
\flushbottom

\section{Introduction}
\label{sec:intro}

The dark photon (DP) - also know as the ``paraphoton'' or ``hidden photon'' - is a hypothetical particle identified as the massive gauge boson associated with a new local $U(1)$ symmetry beyond the standard model (SM) gauge group~\cite{Okun:1982xi}. 
The simplest version of this extension involves assuming all SM particles uncharged with respect to this new symmetry, so that the DP interacts with the SM only through a kinetic mixing term, parameterised by the parameter $\chi$, as shown in the Lagrangian below
\begin{equation}
    \mathcal{L} = -\frac{1}{4} A_{\mu \nu} A^{\mu \nu} - A_\mu j^\mu -\frac{1}{4} B_{\mu \nu} B^{\mu \nu} + \frac{1}{2} m^2 B_\mu B^\mu - \frac{1}{2} \chi A_{\mu \nu} B^{\mu \nu}\,,
\label{eq:lagrangian}
\end{equation}
where $A^\mu$ is the SM photon field, $A^{\mu\nu}$ its field strength tensor, $j^\mu$ is the electromagnetic (EM) 4-current, $B^\mu$ is the DP field with $B^{\mu\nu}$ its field strength tensor, and $m$ is the mass of the DP~\cite{Redondo:2008aa,An:2013yfc}. 
We have assumed here a fixed mass arising from the Stuekelberg mechanism~\cite{stuekelberg-mech}.\footnote{
Another possibility is that the DP gets its mass dynamically through the spontaneous breaking of the $U(1)$ symmetry and the addition of a dark Higgs mechanism \cite{higgs-case}. 
The results for a DP of this type would differ from those found in this paper.
However, since this model is already strongly constrained by isocurvature perturbations \cite{higgs-case-constraints}, at least for the case of DP dark matter, here we will only consider the Stuekelberg case and leave the discussion of the Higgsed case to a future work.}

\begin{figure}
    \centering
    \includegraphics[width=\textwidth,keepaspectratio]{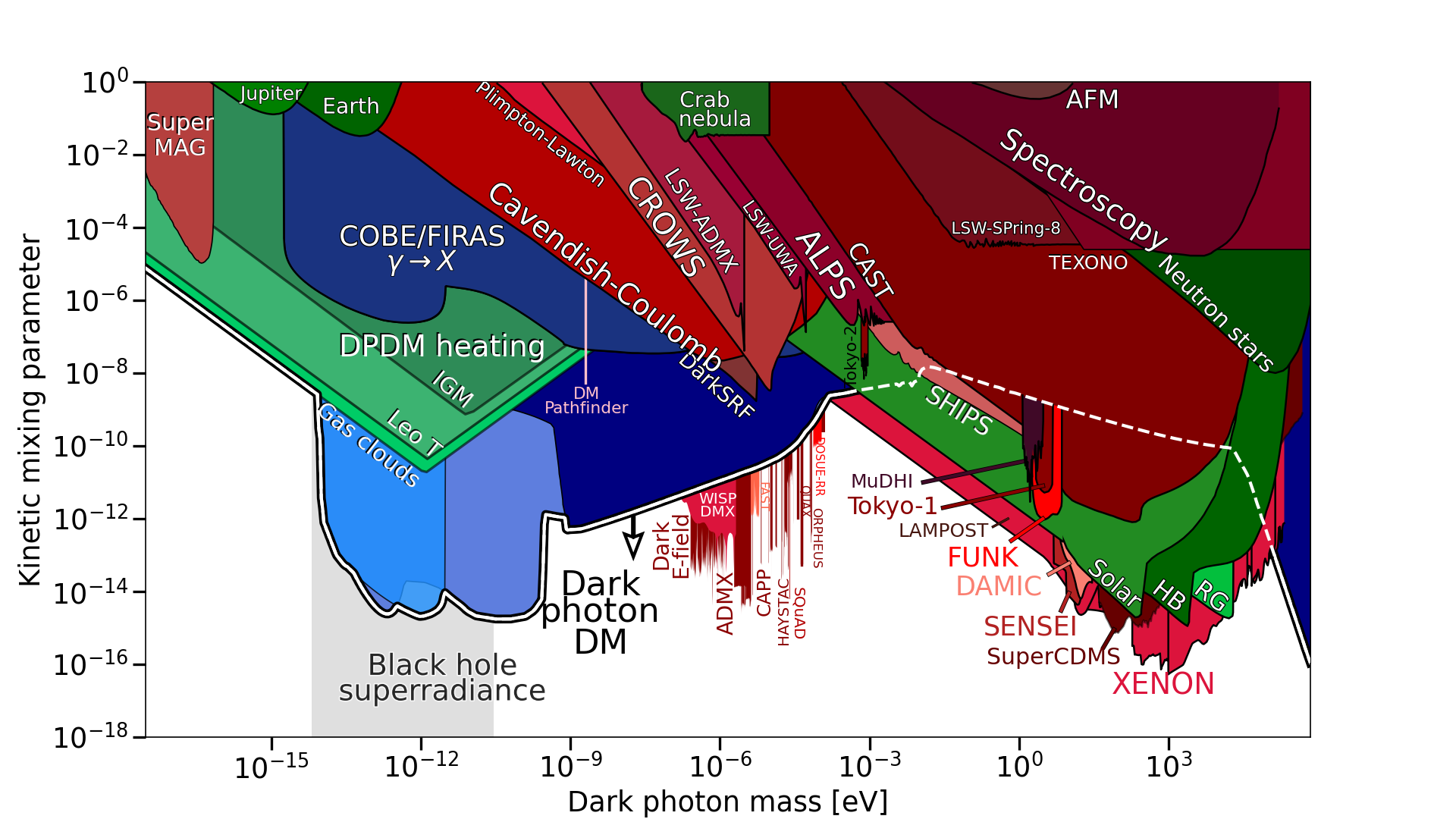}
    \caption{Plot of the current best limits on the dark photon kinetic mixing parameter ($\chi$) and mass ($m$) by various experiments. Plot generated from Ref.~\cite{AxionLimits}.}
    \label{fig:oldLimits}
\end{figure}

The addition of the DP field to the SM leads to a rich phenomenology (see, e.g., Refs.~\cite{Agrawal:2021dbo,Caputo:2021eaa,Antel:2023hkf} for recent reviews).
First of all, the DP is a viable dark matter candidate, which could be produced in the early universe  non-thermally through the misalignment mechanism \cite{DPmisalignment,Graham:2015rva}. 
More generally, the DP could play the role of a vector portal into a dark sector, containing hidden fields. 
There have been a number of studies constraining the DP parameters by inferences from cosmology, astrophysical observations and direct experiments (see, e.g., Ref.~\cite{Caputo:2021eaa} for a recent review).
These constraints are summarised in Fig. \ref{fig:oldLimits}.

In this work, we consider, for the first time, the DP detection potential of the International AXion Observatory (IAXO), a next generation axion helioscope~\cite{IAXO:2019mpb,IAXO:2020wwp}. 
Though designed principally to detect solar axions, IAXO will also be sensitive to other hypothetical particles, including the DP. 
It will consist of a long chamber containing a number of bores inside a strong magnetic field on a movable platform such that it is pointed towards the sun approximately 50\% of the time. 
At the end of the bores are optics that will focus X-rays onto a low-background micromegas detector~\cite{IAXO:2019mpb}. 
Notice that the magnetic field, essential for the conversion of axions into photons, has no effect on the detection of DPs.
The first IAXO iteration will consist of a reduced version, with respect to the original design, called BabyIAXO~\cite{IAXO:2020wwp}. 
The possibility of upgrading to a more powerful setup, called IAXO+, which would greatly enhance the detection possibility, is also contemplated by the IAXO collaboration~\cite{IAXO:2019mpb}. 
The results from the analysis presented in this work will be applied to all 3 scenarios. 
In all cases, the IAXO collaboration plans a first run with a vacuum in the bores, followed by others with the bores filled with $^4$He up to a pressure of $\sim 1$ bar \cite{IAXOconcept}, which would allow resonant conversion of DPs.
In this work, we study the helioscope sensitivity in both configurations.

The paper is organised as follows.
We begin, in Section~\ref{sec:production}, by recapping the derivation of the solar DP flux, following Ref.~\cite{Redondo:2008aa, An:2013yfc, Redondo:2013lna, Redondo:2015iea}, and apply this to the IAXO vacuum run in Section~\ref{sec:detection}. 
In Section~\ref{sec:gas}, we extend the analysis to the IAXO run using a buffer gas. 
In Section~\ref{sec:bfields}, we calculate the contribution of the additional DP flux from the solar magnetic fields, 
while in Section~\ref{sec:pp} we consider the contributions from nuclear processes.
A discussion of the detection of longitudinal DPs is found in Section~\ref{sec:Lpure}.
Finally, in Section~\ref{sec:conc} we present our discussion and conclusions. 
Some technical aspects are left to the appendices. 
The code used to obtain the results shown in this paper can be found at Ref.~\cite{darkphoton_github}.

\section{Thermal dark photon production}
\label{sec:production}
 
In this section, we provide a brief overview of the thermal production of DPs in the solar plasma.  
For more details, the reader is referred to Refs.~\cite{Redondo:2008aa,Redondo:2013lna,Redondo:2015iea}. 
Our notation follows closely that of Ref.~\cite{Redondo:2008aa}.

The Lagrangian in Eq.~\eqref{eq:lagrangian} shows explicitly that the $A^{\mu}$ and the $B^\mu$ fields are non-orthogonal because of the kinetic mixing, parameterised by the coupling  $\chi$. 
 This parameter is limited to $\chi\ll1$ by phenomenological considerations and experimental constraints, as shown in Fig.~\ref{fig:oldLimits}.
In fact, it will be shown below that $\chi \lesssim 10^{-10}$ in the region of interest of this analysis.
%
Following Ref.~\cite{Redondo:2008aa}, we conveniently remove the kinetic mixing term by introducing the \emph{sterile} state $S^\mu$, orthogonal to the photon field
\begin{equation}
S^\mu \equiv B^\mu + \chi A^\mu \,.
\label{eq:sterile}
\end{equation}
The Lagrangian in Eq.~(\ref{eq:lagrangian}) can then be rewritten in terms of $S^\mu$
\begin{equation}
    \mathcal{L} = -\frac{1}{4} \left( (1-\chi^2)A_{\mu\nu}A^{\mu\nu} + S_{\mu\nu}S^{\mu\nu} \right) - A_\mu J^\mu + \frac{1}{2} m^2 (S_\mu - \chi A_\mu)(S^\mu - \chi A^\mu),
    \label{eq:lagrangian2}
\end{equation}
and by the field redefinitions
\begin{subequations}
\begin{equation}
    A^\mu \to \frac{1}{\sqrt{1-\chi^2}} A^\mu \,,
\end{equation}
\begin{equation}
    \chi \to \sqrt{1-\chi^2} \chi \,,
\end{equation}
\begin{equation}
    J^\mu \to \sqrt{1-\chi^2} J^\mu \,,
\end{equation}
\end{subequations}
where the redefinition of $J^\mu$ is equivalent to redefining the electric charge $e \to \sqrt{1-\chi^2} e$, we can express Eq.~\eqref{eq:lagrangian2} as
\begin{equation}
    \mathcal{L} = -\frac{1}{4} \left(A_{\mu\nu}A^{\mu\nu} + S_{\mu\nu}S^{\mu\nu} \right) - A_\mu J^\mu + \frac{1}{2} m^2 (S_\mu - \chi A_\mu)(S^\mu - \chi A^\mu)\,.
    \label{eq:lagrangian3}
\end{equation}
and we get equations of motion (EoM) in the Lorenz gauge $\partial_\mu A^\mu = 0\,,$\footnote{
 Note that the Lorentz gauge for the hidden photon field $\partial_\mu B^\mu = 0$ has been assumed from the start since Eq.~\ref{eq:lagrangian} is gauge fixed in such a way that the Stuekelberg scalar does not appear. The condition $\partial_\mu S^\mu = 0$ follows from the gauge fixing of the $A^\mu$ and $B^\mu$ fields.}
\begin{equation}
\left( (1 - \chi^2m^2)K^2 g^{\mu \nu} - \frac{1}{2} (\Pi^{\mu \nu} + \Pi^{\nu \mu}) \right)A_\nu + \chi m^2 S^\mu = 0,
\label{eq:EoM1}
\end{equation}
\begin{equation}
\left( K^2 - m^2 \right)S^\mu + \chi m^2 A^\mu = 0,
\label{eq:EoM2}
\end{equation}
where $K^\mu=(\omega,\bm{k})$ is the 4-momentum of the particle, $g^{\mu \nu}$ is the metric,\footnote{In this paper we assume the convention $g^{\mu\nu} = \eta^{\mu\nu} \equiv \operatorname{diag}(1,-1,-1,-1)$ for the flat space (Minkowski) metric, and use natural units defined as $\hbar = c = k_B = 1$, such that the angular frequency $\omega$ corresponds to the energy and the wavenumber $\bm{k}$ to the momentum.} and $\Pi^{\mu\nu}$ is the polarisation tensor in the plasma, defined according to $ A_\mu J^\mu = -\frac{1}{2} A_\mu \Pi^{\mu\nu} A_\nu$~\cite{melroseCovariantDispersion}. 
It is clear from Eq.~\ref{eq:EoM1} that the antisymmetric component of $\Pi^{\mu\nu}$ cancels out. 
In an isotropic medium $\Pi^{\mu\nu}$ is entirely symmetric so we see $\frac{1}{2} (\Pi^{\mu \nu} + \Pi^{\nu \mu}) = \Pi^{\mu\nu}$.
Since $\chi \ll 1$, the $\chi^2$ terms can be safely ignored and Eq.~(\ref{eq:EoM1}) simplifies to
\begin{equation}
\left( K^2 g^{\mu \nu} - \Pi^{\mu \nu} \right)A_\nu + \chi m^2 S^\mu = 0.
\label{eq:EoM1-simplified}
\end{equation}
We will continue to keep only the lowest order terms in $\chi$ in the rest of this paper. 
Without loss of generality, we assume that the wave propagates along the $z$-axis. 
In this reference frame, we can define a set of orthonormal polarisation vectors as
\begin{subequations}
\label{eq:basisvectors}
\begin{equation}
e_l^\mu = \sqrt{\frac{1}{\omega^2 - k^2}} (k,0,0,\omega)\,,
\label{eq:el}
\end{equation}
\begin{equation}
e_x^\mu = (0,1,0,0)\,,
\label{eq:ex}
\end{equation}
\begin{equation}
e_y^\mu = (0,0,1,0)\,,
\label{eq:ey}
\end{equation}
\end{subequations}
where $k \equiv |\bm{k}|$.
These vectors satisfy the conditions $e_a^\mu e_{b,\mu} = -\delta_{ab}$ and $e_a^\mu K_\mu = 0$.  
Furthermore, the conservation of current and gauge invariance require that $K_\mu \Pi^{\mu \nu} = \Pi^{\mu \nu} K_\nu = 0$ (see, e.g., Ref.~\cite{Raffelt:1996wa}).
As the vectors $e_a^\mu$ together with $K^\mu$ form a complete basis and the condition $e_a^\mu K_\mu = 0$ is satisfied, the most general $\Pi^{\mu \nu}$ can be constructed as $\Pi^{\mu \nu} = \sum_{a,b} - \pi_{ab} e_a^\mu e_b^\nu$.
However, assuming azimuthal symmetry around the $k$-direction in the solar plasma, the only non-zero terms are those containing $e_a^\mu e_a^\nu$ and thus the polarisation tensor 
reduces to
\begin{equation}
\Pi^{\mu \nu} = \sum_{a} - \pi_{a} e_a^\mu e_a^\nu
\label{eq:piDecompSym}
\end{equation}
where we have defined $\pi_a \equiv \pi_{aa}$. Inserting Eq.~\eqref{eq:piDecompSym} into Eq.~\eqref{eq:EoM1-simplified}, contracting Eq.~\eqref{eq:EoM1-simplified} and Eq.~\eqref{eq:EoM2} with $e_a^\mu$, and defining $A_a \equiv e_a^\mu A_\mu$ and $S_a \equiv e_a^\mu S_\mu$ returns
\begin{subequations}
\label{eq:EoMs}
\begin{equation}
\left( K^2 - \pi_a \right)A_a + \chi m^2 S_a = 0
\label{eq:EoM3}
\end{equation}
\begin{equation}
\left(K^2 - m^2 \right)S_a + \chi m^2 A_a = 0 .
\label{eq:EoM4}
\end{equation}
\end{subequations}
Notice that the EoM of different polarisations are all independent. 
Because of isotropy, the $x$ and $y$ polarised plasmons can be treated equivalently and will henceforth commonly be labeled transverse or T-plasmons, as opposed to longitudinal or L-plasmons.

\begin{figure}
    \centering
    \includegraphics[width=\textwidth,keepaspectratio]{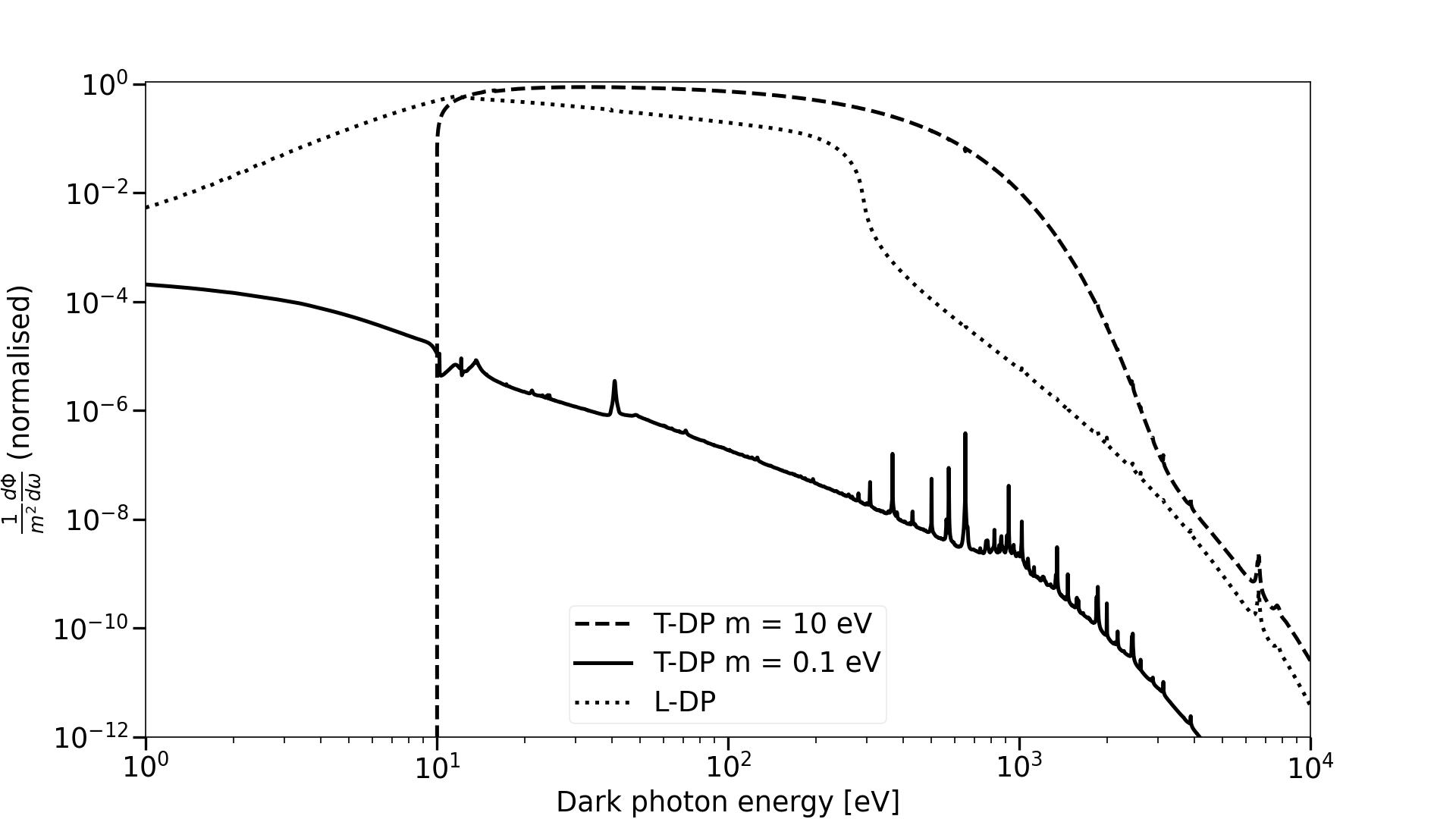}
        \caption{The differential flux spectra $(d\Phi/d\omega) \, /m^2$ for various thermal dark photon production, normalised to the maximum of the $m=10$ eV line, $(d\Phi/d\omega) \, /(\chi^2m^2) = 1.01 \times 10^{36}$ cm$^{-2}$ s$^{-1}$ eV$^{-3}$.}
    \label{fig:spectrum}
\end{figure}

The information about the dynamics is contained in the photon self-energy. In particular, the real part gives the effective mass of each polarisation, while the imaginary part is related to the production and absorption in the medium~\cite{Weldon:1983jn,Redondo:2013lna,Redondo:2015iea}
\begin{subequations}
\label{eq:ReImpi}
\begin{equation}
\operatorname{Re}(\pi_t) = \frac{\omega^2}{K^2} \operatorname{Re}(\pi_l) \equiv m_{\gamma}^2\,,
\label{eq:Repi}
\end{equation}
\begin{equation}
\operatorname{Im}(\pi_t) = \frac{\omega^2}{K^2} \operatorname{Im}(\pi_l) \equiv - \omega \Gamma \, ,
\label{eq:Impi1}
\end{equation}
\begin{equation}
\Gamma = (e^{\omega / T} - 1) \Gamma^{\operatorname{prod}} = (1-e^{-\omega / T}) \Gamma^{\operatorname{abs}} \,,
\label{eq:Impi2}
\end{equation}
\end{subequations}
where $T$ is the temperature. Here $\Gamma^{\operatorname{prod}}$ and $\Gamma^{\operatorname{abs}}$ are the rates for production and absorption respectively.

In the non-relativistic, non-degenerate limit, the effective photon mass $m_\gamma^2$, given by the plasma frequency $\omega_p$, is approximated as
\begin{equation}
    m_\gamma^2 = \omega_{\mathrm{p}}^2 \approx \frac{4\pi\alpha n_e}{m_e}\,,
    \label{eq:plasmaFreq}
\end{equation}
where $\alpha$ is the fine structure constant, $n_e$ is the electron number density and $m_e$ is the electron mass.
The photon absorption at high energies is dominated by inverse bremsstrahlung (free-free absorption) and Thomson scattering. 
At low energies, bound-free and bound-bound processes become important. 
As we will consider DP energies down to $\sim 1$ eV, it will be necessary to include all of these processes. 
Refer to Appendix~\ref{app:Gamma} for more details.

Eqs.~\eqref{eq:EoMs} imply the oscillations between the SM photon field $A_a$ and the sterile DP field $S_a$ for a given polarisation $a$. 
In order to calculate the flux of transverse DPs, we need to identify the propagation states $\tilde{A_a}$ and $\tilde{S_a}$ that diagonalise the EoM
\begin{subequations}
\label{eq:propStates}
\begin{equation}
    \tilde{A_a} = A_a - \frac{\chi_a m^2}{\pi_a - m^2} S_a
    \label{eq:propStateA}
\end{equation}
\begin{equation}
    \tilde{S_a} = S_a + \frac{\chi_a m^2}{\pi_a - m^2} A_a
    \label{eq:propStateS},
\end{equation}
\end{subequations}
where $\chi_l \equiv \frac{\omega}{m} \chi$ and $\chi_t \equiv \chi$ to account for the renormalisation of $\chi$ required in the L-plasmon case~\cite{Redondo:2013lna}.
We note that the interactions of the plasma produce pure photon states $A_a$, which are a combination of photon-like states $\tilde A_a$ and sterile-like states $\tilde S_a$. 
Both are absorbed by the medium by virtue of their photonic component, but in the small-mixing angle case considered here the photon-like component is absorbed on one photonic mean-free-path, $\lambda\sim \Gamma^{-1}$, while the sterile-like component has a much longer absorption length $\propto \chi^{-2}\Gamma^{-1}$.   
Thus, we find the conversion probability from the initial photon state to the propagating sterile state
\begin{subequations}
\label{eq:l-tProbs}
\begin{equation}
    P_{A_t \rightarrow \tilde S_t} = \frac{\chi^2 m^4}{(m_{\gamma}^2 - m^2)^2 + (\omega \Gamma)^2}\,,
    \label{eq:tProb}
\end{equation}

\begin{equation}
    P_{A_l \rightarrow \tilde S_l} = \frac{\chi^2 m^2 \omega^2}{(m_{\gamma}^2 - \omega^2)^2 + (\omega \Gamma)^2}\,.
    \label{eq:lProb}
\end{equation}

\end{subequations}

The value of the photon-sterile photon mixing angle changes with the properties of the plasma as the DP propagates outwards but the sterile-like DPs follows an effectively adiabatic exit without changing much its amplitude as long as the mixing angle is always small. Naturally, the sterile-like propagation eigenstate at zero-density ($\Pi_{\mu\nu}=0$), i.e. upon exiting the Sun, is the original $B$ state.  

Multiplying the conversion probabilities in Eq. \eqref{eq:l-tProbs} by the rate of plasmon creation and integrating over the solar interior, we obtain the flux of $\tilde{S_a}$ DPs on Earth~\cite{Redondo:2015iea}:
\begin{equation}
    \frac{d\Phi_t}{d\omega} = \chi^2 m^4 \int_0^{R_\odot} \frac{r^2 dr}{\pi^2 d_\odot^2} \frac{\omega \sqrt{\omega^2 - m^2}}{(m^2 - m_\gamma^2)^2 + (\omega \Gamma)^2} \frac{\Gamma}{e^{\omega/T} - 1}\,,
    \label{eq:phiT}
\end{equation}
and
\begin{equation}
    \frac{d\Phi_l}{d\omega} = \chi^2 m^2 \omega^2 \int_0^{R_\odot} \frac{r^2 dr}{2 \pi^2 d_\odot^2} \frac{\omega \sqrt{\omega^2 - m^2}}{(\omega^2 - m_\gamma^2)^2 + (\omega \Gamma)^2} \frac{\Gamma}{e^{\omega/T} - 1} \,,
    \label{eq:phiL}
\end{equation}
where $d_\odot$ is the mean earth-sun distance and $R_\odot$ is the solar radius.
Notice that the T-DP flux is suppressed by a factor of $m^2/\omega^2$ with respect to the L-DP flux and has a resonance at $m^2 \approx \omega_{\operatorname{p}}^2$ whereas the L-DP is resonantly produced at $\omega \approx \omega_{\operatorname{p}}$, and that the T-DP flux has been multiplied by 2 to account for the 2 polarisations. 
These differences and their implications are discussed further in Section~\ref{sec:bfields} and \ref{sec:Lpure}. 
The differential flux spectra $\frac{d\phi}{d\omega}$ as a function of $\omega$ for both L- and T- DPs are presented in Fig.~\ref{fig:spectrum} for various values of $m$. 

\section{Detection by IAXO}
\label{sec:detection}

In this section we estimate the IAXO sensitivity to the DP solar flux presented above. 
Since only transverse modes can convert into regular photons in a vacuum and IAXO will initially operate with vacuum in its bores,
in this section we will ignore the longitudinal mode and focus only on the flux in Eq.~\eqref{eq:phiT}. 
In this case, the propagating mode is the $\tilde{S_t}$ state, which is a superposition of regular and sterile photon states as shown in Eq.~\eqref{eq:propStateS}. 
However, upon interaction with the IAXO shielding, the photon component $A_t$ will be absorbed leaving only the pure sterile state $S_t$.
Thus, to estimate the IAXO sensitivity we need to calculate the probability of $S_t \rightarrow A_t$ in the IAXO bores and integrate over the range of energies detectable by IAXO. 
The conversion probability over the length $L$ can be calculated 
starting from the propagation equations~\eqref{eq:EoMs} and the definition of the propagation states Eq.~\eqref{eq:propStates}, following a general strategy outlined, e.g., in Ref.~\cite{Redondo:2008aa,Redondo:2013lna}. We find, for T-DPs
\begin{equation}
    P_{S_t \rightarrow A_t} = \frac{\chi^2 m^4}{(m_{\gamma}^2 - m^2)^2 + (\omega \Gamma)^2} \left( 1 + e^{-\Gamma L} - 2e^{-\Gamma L / 2} \cos(q_t L) \right), 
    \label{eq:conversionProb}
\end{equation}
where $m_\gamma$ and $\Gamma$ refer to the effective photon mass and absorption length in the IAXO bores. 
Assuming the interior of the IAXO bores is a vacuum, $m_\gamma = \Gamma = 0$, Eq.~\eqref{eq:conversionProb} for T-plasmons reduces to
\begin{equation}
    P_{S_t \rightarrow A_t} = 4 \chi^2 \sin^2\left(\frac{q_t L}{2} \right)\,,
    \label{eq:Tconversion}
\end{equation}
where $q_t = \sqrt{\omega^2 - m_{\gamma}^2} - \sqrt{\omega^2 - m^2}$. The length $L$ is defined as the distance between the shielding and the optics. 

Multiplying the differential flux \eqref{eq:phiT} by the conversion probability \eqref{eq:Tconversion} and integrating over the range of detectable energies gives the flux of T-DPs detected by IAXO. The detector parameters have been taken from Ref.~\cite{IAXO:2019mpb}.
We have also assumed that the whole sun is visible to the optics with a uniform optics efficiency $\epsilon_o$ as defined in \cite{IAXO:2019mpb}, and a uniform detector efficiency $\epsilon_d$ over the energy range considered. The implementation of specific detector and optical technologies is deferred to a future work.

\begin{figure}
    \centering
    \includegraphics[width=\textwidth,keepaspectratio]{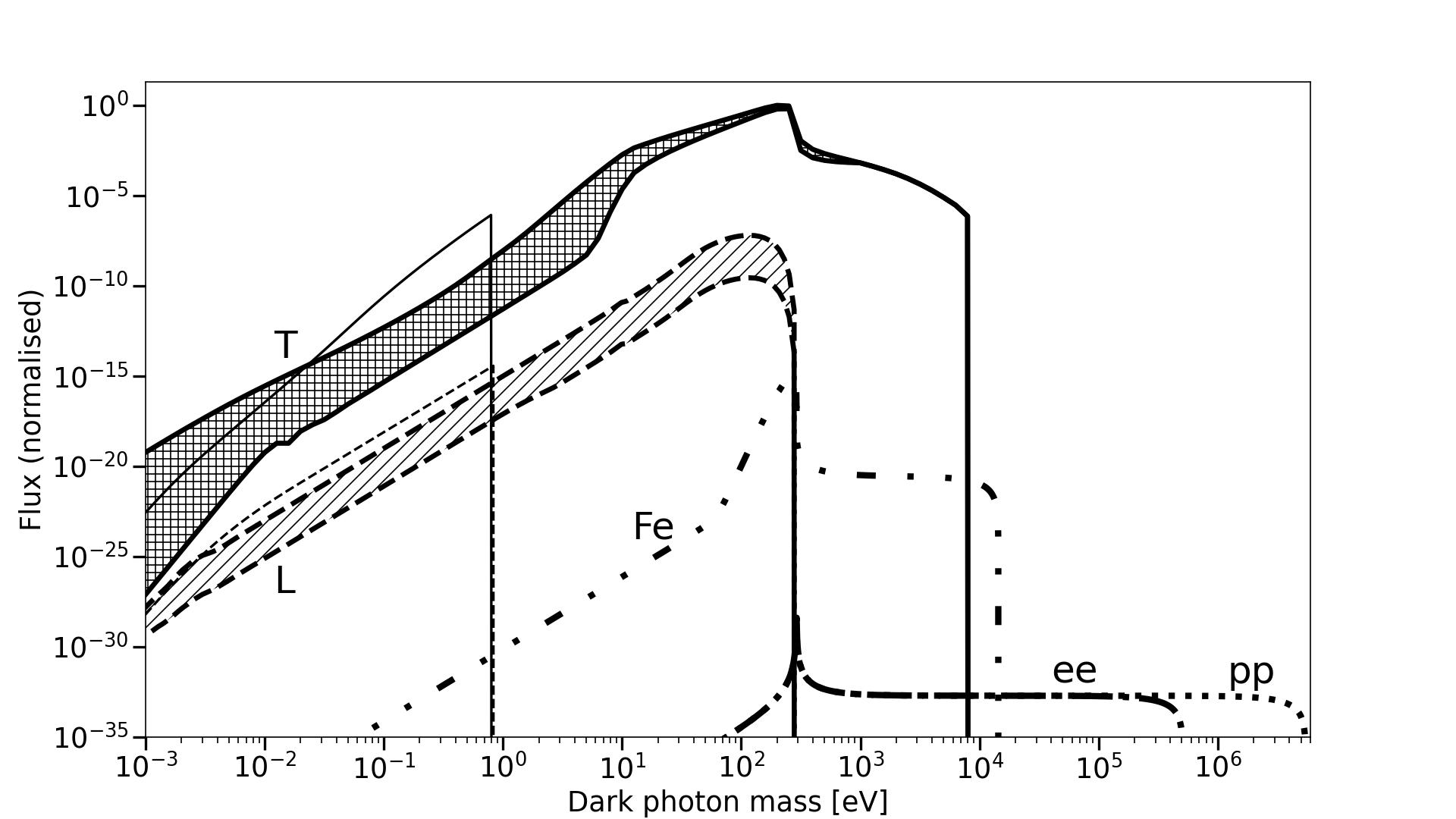}
    \caption{Comparison of the total back-converted photon flux arriving at the BabyIAXO detector as a function of the dark photon mass for the various dark photon production and conversion mechanisms discussed in this paper. 
    The `T' lines represent the flux from standard thermal production, with the thick lines corresponding to the vacuum run and thin lines to the gas run. The cross hatched section shows the potential range of fluxes depending on the lower threshold of detectable photon energies, with the lower line representing a threshold of 1 keV and the upper of 1 eV. For the gas run a threshold of 30 eV is assumed.    
    The `L' lines show the flux of T-DPs emitted from L-plasmons (see Sec.~\ref{sec:bfields}), again with the thick line corresponding to the BabyIAXO vacuum run with an energy threshold of 1 eV, and the thin line corresponding to the gas run with a threshold of 30 eV. All of these assume an upper detection energy of 10 keV. The shaded region between the `L' vacuum run lines shows the effect of the uncertainty on the solar magnetic field strength (see Sec.~\ref{sec:bfields} for details). The same band would be present under the gas run curve but is omitted for readability.
    The non-thermal contributions - `pp' for the 5.49 MeV DPs, `ee' for e+ e- annihilation, and `Fe' for the $^{57}$Fe gamma transition - are all emitted with distinct peak energies so it is assumed that BabyIAXO is sensitive to the relevant energies in each case. The flux is normalised to the peak value of the `T' vacuum run curve, $\Phi/\chi^4 = 3.07 \times 10^{43}$ cm$^{-2}$ s$^{-1}$.}
    \label{fig:comparison}
\end{figure}

Notice that the detector's lower energy threshold has a significant impact on the detection sensitivity. 
The spectra in Fig.~\ref{fig:spectrum} make it clear that most of the solar DPs have $\omega < 1$ keV so a detector that is not sensitive to these energies would miss the majority of the DP flux. 
The cross-hatched region marked `T' in Fig.~\ref{fig:comparison} shows how the detected flux is affected by the energy threshold, with the lower line showing the flux for a threshold of 1 keV and the upper for 1 eV. 
It is evident from the figure that for $m \lesssim 100$ eV, the DP flux visible to IAXO is greatly increased by lowering the threshold.

In Fig.~\ref{fig:limits}, we show the 95\% confidence level (CL) limits IAXO would set in case of no signal events.
In deriving these bounds, we used the maximum likelihood method of Ref.~\cite{juliaThesis}, described in Appendix~\ref{app:likelihood}. 
Comparing our results to the limits shown in Fig.~\ref{fig:oldLimits} reveals that IAXO would improve substantially over CAST and may surpass solar energy loss bounds. 
However, the IAXO limits would be less stringent than the limits set by the XENON experiment~\cite{xenon}.

\section{Resonant conversion using $^4$He}
\label{sec:gas}

As mentioned before, the IAXO bores can be filled with $^4$He to stimulate the resonant conversion of DPs into SM photons. 
As the energies considered are far greater than the excitation energies of the gas, we can assume that photons travel as in a plasma and define a plasma frequency for the buffer gas as in Eq.~\eqref{eq:plasmaFreq} \cite{Redondo:2008aa}. Due to its design, IAXO can support pressures that give $m_\gamma \le 1$ eV~\cite{IAXO:2019mpb}.
Resonant conversion of T-DPs occurs at $m_\gamma = m$ which implies $\Delta p = 0$ and $\Gamma \not= 0$, so that Eq.~\eqref{eq:conversionProb} reduces to 
\begin{equation}
    P_{S_t \rightarrow A_t} = \frac{\chi^2 m^4}{(\omega \Gamma)^2} \left( 1 + e^{-\Gamma L} - 2e^{-\Gamma L / 2} \right)  \,.
    \label{eq:gasConversion}
\end{equation}
In the limit $\Gamma_t L \rightarrow 0$, this can be approximated as
\begin{equation}
    P_{S_t \rightarrow A_t} \approx \frac{\chi^2 m^4 L^2}{4 \omega^2}.
    \label{eq:gasConversionApprox}
\end{equation}
The absorption length $\Gamma$ for photons in the $^4$He buffer gas can be calculated 
following the strategy outlined in Ref.~\cite{juliaThesis} and in Appendix \ref{app:buffergas}.

Replacing Eq.~\eqref{eq:Tconversion} with Eq.~\eqref{eq:gasConversion} in the integral, one can calculate the flux of back-converted photons detected by IAXO in the presence of a buffer gas. As mentioned in Section~\ref{sec:detection}, the range of detectable energies affects the flux seen by IAXO. However the approximations we have used for the photon absorption and effective mass are invalid for photon energies below the ionisation energy of $^4$He where atomic structure becomes relevant. For this reason we have assumed a threshold of 30 eV for the buffer gas run.

\begin{figure}
    \centering    \includegraphics[width=\textwidth,keepaspectratio]{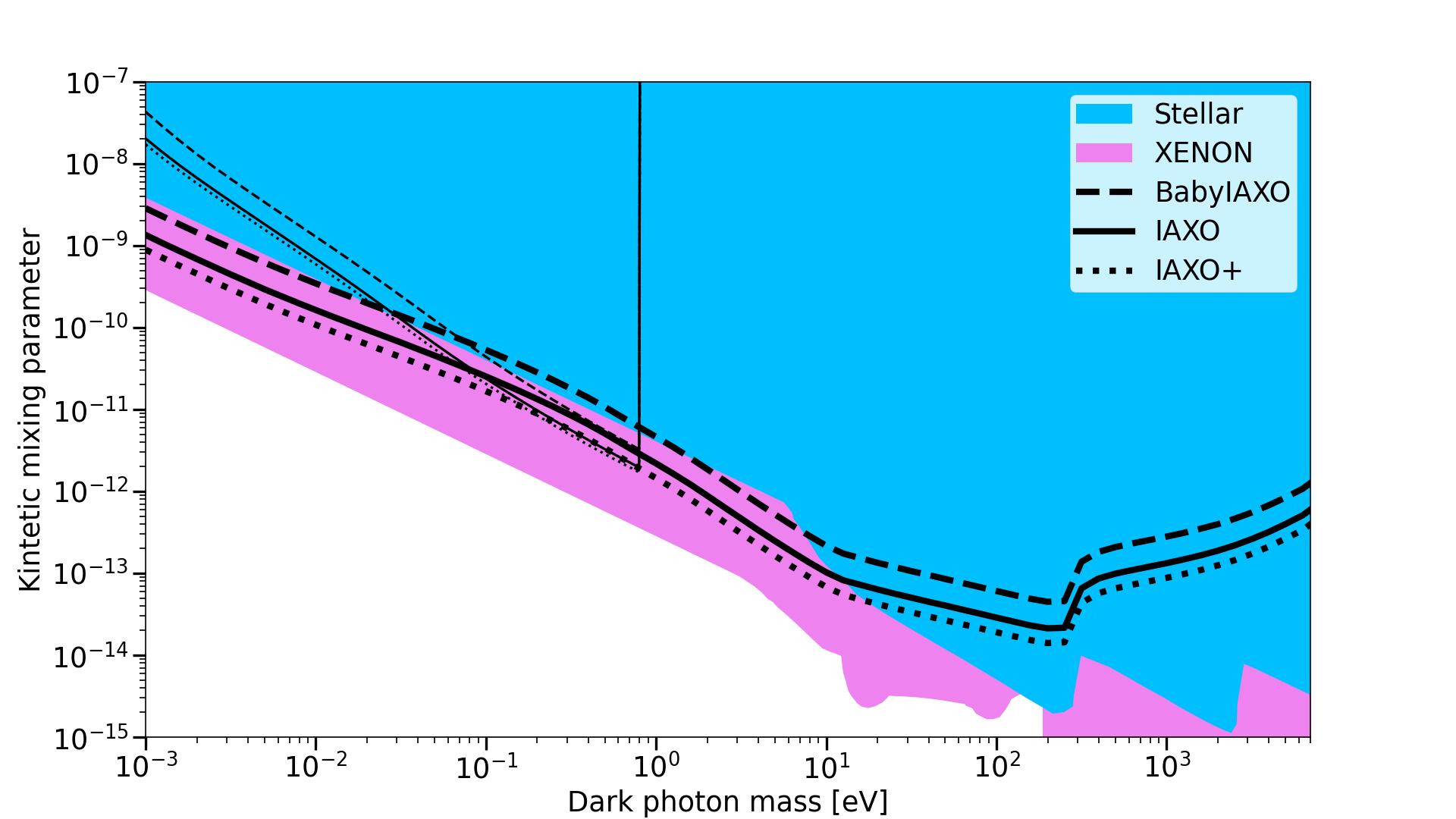}
        \caption{The 95\% CL limits that would be placed by IAXO on the DP parameter space by the method presented in Appendix~\ref{app:likelihood}, assuming no discovery. The thick lines correspond to the vacuum runs, and the thin lines to the buffer gas runs.}
    \label{fig:limits}
\end{figure}

The detected flux from the gas run is plotted as the thin `T' line in Fig.~\ref{fig:comparison}, where it has been assumed that the gas pressure has been increased in steps to a maximum of 10 bar to give resonant conversion for a range of DP masses, with 5 days of data taking on each pressure step and the statistics from Appendix~\ref{app:likelihood}. The limits calculated here and in Section~\ref{sec:detection} are shown along with the stellar and XENON bounds in Fig.~\ref{fig:limits}.

\section{Contribution of solar magnetic fields}
\label{sec:bfields}

So far we have assumed the solar plasma to be isotropic. 
However the sun houses large magnetic fields, including potentially very large fields in its core, that cause plasma anisotropies. 
This anisotropy implies that the plasmon propagation states are superpositions of L and T polarisiation states, and thus  introduces a non-vanishing probability of an L-plasmon oscillating into a T-plasmon. 
Such oscillations provide a potentially quite interesting T-plasmon production channel.
In  fact, the production of L-plasmons is enhanced by a factor of $\omega^2/m^2$ with respect to T-plasmons, a factor which could be quite large at low masses.
Furthermore, the L modes are resonantly produced throughout the whole sun, rather than in a specific shell, as detailed in Section~\ref{sec:detection}. 
Helioscopes, at least in their basic configuration, cannot detect L-plasmons. 
Thus, the possibility that a fraction of the more efficiently produced L-modes might convert into  T-modes in the anisotropic plasma is worth considering. 

In an anisotropic plasma  
the polarisation tensor cannot be reduced to the diagonal form given in Eq.~\eqref{eq:piDecompSym} but must be kept in a general form. 
The resulting EoM are
\begin{equation}
\left( K^2 - \pi_a \right)A_a - \sum_{a \not = b} \pi_{ab} A_b + \chi m^2 S_a = 0
\label{eq:EoM5}
\end{equation}
and so the L, $x$ and $y$ plasmons are all coupled.

The plasmon mixing term $\pi_{ab}$ can be expressed in terms of the dielectric tensor $\epsilon^{ij}$ in a magnetised plasma using the following relation (Cf. Ref.~\cite{melroseCovariantDispersion})
\begin{equation}
\Pi^{ij} = \omega^2 (\epsilon^{ij} - \delta^{ij}) .
\label{eq:dielectricConversion}
\end{equation}
The components of the dielectric tensor can be found in Ref.~\cite{ryluk} assuming an ideal, non-relativistic plasma.\footnote{Note that in Ref.~\cite{ryluk}, $\bm{B}$ is along the $z$-axis, 
so it must be rotated to give the form used here where $\bm{k}$ is along the $z$-axis and $\bm{B}$ lies in the $x-z$ plane.}
Then, the components of the polarisation tensor arising from the anisotropy can be put in the form
\begin{equation}
\Pi^{ij}_B = - \omega_{\operatorname{p}}^2 \left( \begin{array}{ccc}
1 & 0 & -g^2 \sin\theta \cos\theta \\
0 & 1 & 0 \\
-g^2 \sin\theta \cos\theta & 0 & 1
\label{eq:dielectic}
\end{array} \right)
\end{equation}
where
\begin{equation}
g \equiv \frac{\omega_B}{\omega},
\label{eq:g}
\end{equation}
with $\omega_B \equiv e B / m_e$ the electron cyclotron frequency, and 
$\theta$ the angle between the direction of propagation of the wave and the magnetic field. 
Note that according to Ref.~\cite{ryluk} there is a purely imaginary antisymmetric component to $\Pi^{ij}_B$ at $\mathcal{O}(g)$, 
but as discussed in Section~\ref{sec:production}, this component does not play a role in the equations of motion 
and we are left with the lowest order symmetric term in Eq.~\eqref{eq:dielectic}.
The above result is valid in the limit $\omega_B \ll \omega$, a limit which is justified in our analysis since $\omega_B \lesssim$ 1 eV even for the largest values of $B$, and we are interested in energies $\omega \gtrsim 1$ eV.
Propagation states can be defined similarly to what done in Section \ref{sec:production}. 
In the limits $\omega_B \ll \omega$ and $m \ll \omega$ we can find the probability of an L-plasmon created in the sun being emitted as a T-DP from the sun's surface
\begin{equation}
    P_{A_l \rightarrow S_t} = \left|\frac{\chi m \omega \pi_{lt}}{\pi_t \left( \pi_t - \omega^2 \right)} \right|^2
    \label{eq:prob1}
\end{equation}
where 
\begin{equation}
    \pi_{lt} = g^2 m_{\gamma}^2 \frac{m}{\omega} \sin{\theta}\cos{\theta}.
    \label{eq:pilt}
\end{equation}
From the point of view of our analysis, this non-vanishing probability of converting an L into a T mode is the most remarkable effect of the magnetic field. However, Eq.~\eqref{eq:pilt} shows that the enhanced conversion probability of L-plasmons compared to T-plasmons by a factor of $\omega^2/m^2$ is suppressed by exactly the same factor in the mixing term $|\pi_{lt}|^2$.
Assuming that the contribution from $\Pi^{ij}_B$ is small compared to that from Eq.~\eqref{eq:mGammaFull} and Eq.~\eqref{eq:GammaFull}, we can use the same forms for $\pi_t, \pi_l$ as in Section~\ref{sec:production} to find
\begin{equation}
    P_{A_l \rightarrow S_t} = \frac{\chi^2 m^4 g^4 m_{\gamma}^4 \sin^2(\theta)\cos^2(\theta)}{(m_{\gamma}^4 + (\omega \Gamma)^2) ((\omega^2 - m_{\gamma}^2)^2 + (\omega \Gamma)^2)} .
    \label{eq:prob2}
\end{equation}
Following again the procedure of Section \ref{sec:production}, and averaging over $\theta$ such that $ \sin^2(\theta)\cos^2(\theta) \rightarrow 1/8$, we obtain an expression for the flux of DPs produced in such a way arriving on Earth
\begin{equation}
    \frac{d\Phi}{d\omega} \approx \chi^2 m^4 \int_0^{R_\odot} \frac{r^2 dr}{16 \pi^2 d_\odot^2} \frac{\omega \sqrt{\omega^2 - m^2} m_{\gamma}^4 g^4}{(m_{\gamma}^4 + (\omega \Gamma)^2) ((\omega^2 - m_{\gamma}^2)^2 + (\omega \Gamma)^2)} \frac{\Gamma}{e^{\omega/T} - 1}.
    \label{eq:lmixing}
\end{equation}

The mixed DP production has a kind of resonance at both $\omega = m_{\gamma}$ and $m = m_{\gamma}$ as is seen in Eq.~\eqref{eq:lmixing}. ``Resonant'' production can be studied by only considering DPs with $\omega = m_{\gamma}$ somewhere in the sun. Using the result from Ref.~\cite{Redondo:2013lna}
\begin{equation}
    \frac{\omega^2 \Gamma}{(\omega^2 - m_{\gamma}^2)^2 + (\omega \Gamma)^2} \sim \frac{\pi}{2} \delta(\omega - m_{\gamma})
    \label{eq:deltafunction}
\end{equation}
which is valid for the weak damping condition $\Gamma \ll m_{\gamma}$, Eq.~\eqref{eq:lmixing} at resonance reduces to
\begin{equation}
    \Phi \approx \chi^2 m^4 \int_0^{R_\odot} \frac{r^2 dr}{32 \pi d_\odot^2} \frac{\omega_B^4}{ m_{\gamma}^4 } \frac{1}{e^{m_{\gamma}/T} - 1}\,,
    \label{eq:lmixingres}
\end{equation}
where we have also assumed $m_{\gamma} \gg m$. Note that the same form for Eq.~\eqref{eq:lmixingres} is obtained by a thermal field theory treatment similar to that outlined in Ref.~\cite{Redondo:2013lna}.

The solar magnetic field model used in our analysis was taken from Ref.~\cite{Bfields} and assumes 3 distinct sections of quadripolar solar magnetic fields all of the form
\begin{equation}
    \bm{B}(r,\theta) = 3 a(r) \cos(\theta) \sin(\theta) \bm{e}_\phi
\end{equation}
where $a(r)$ is a function defined for each solar magnetic field region.

In the radiative zone, which was defined as the region $0 < r \leq r_0$ with $r_0 = 0.712 R_\odot$, the function $a(r)$ was defined as
\begin{equation}
    a(r) = K_\lambda \left( \frac{r}{r_0} \right)^2 \left( 1 - \left( \frac{r}{r_0} \right)^2 \right)^\lambda
\end{equation}
where $K_\lambda \equiv (1 + \lambda) (1 + \lambda^{-1})^\lambda B_0$ and $\lambda \equiv 10 (r_0 / R_\odot) + 1$. 
$B_0$ is the maximum value of the magnetic field. 
The exact value of $B_0$ is unknown.
The range adopted in this work is $B_0\in [200,3000]$T (see Ref.~\cite{Bfields,Hoof:2021mld}). 
This corresponds to the values used also in the recent analyses of the axion production in the solar magnetic field~\cite{Hoof:2021mld, Guarini:2020hps, Caputo:2020quz, OHare:2020wum}.

\begin{figure}
    \centering
    \includegraphics[width=\textwidth,keepaspectratio]{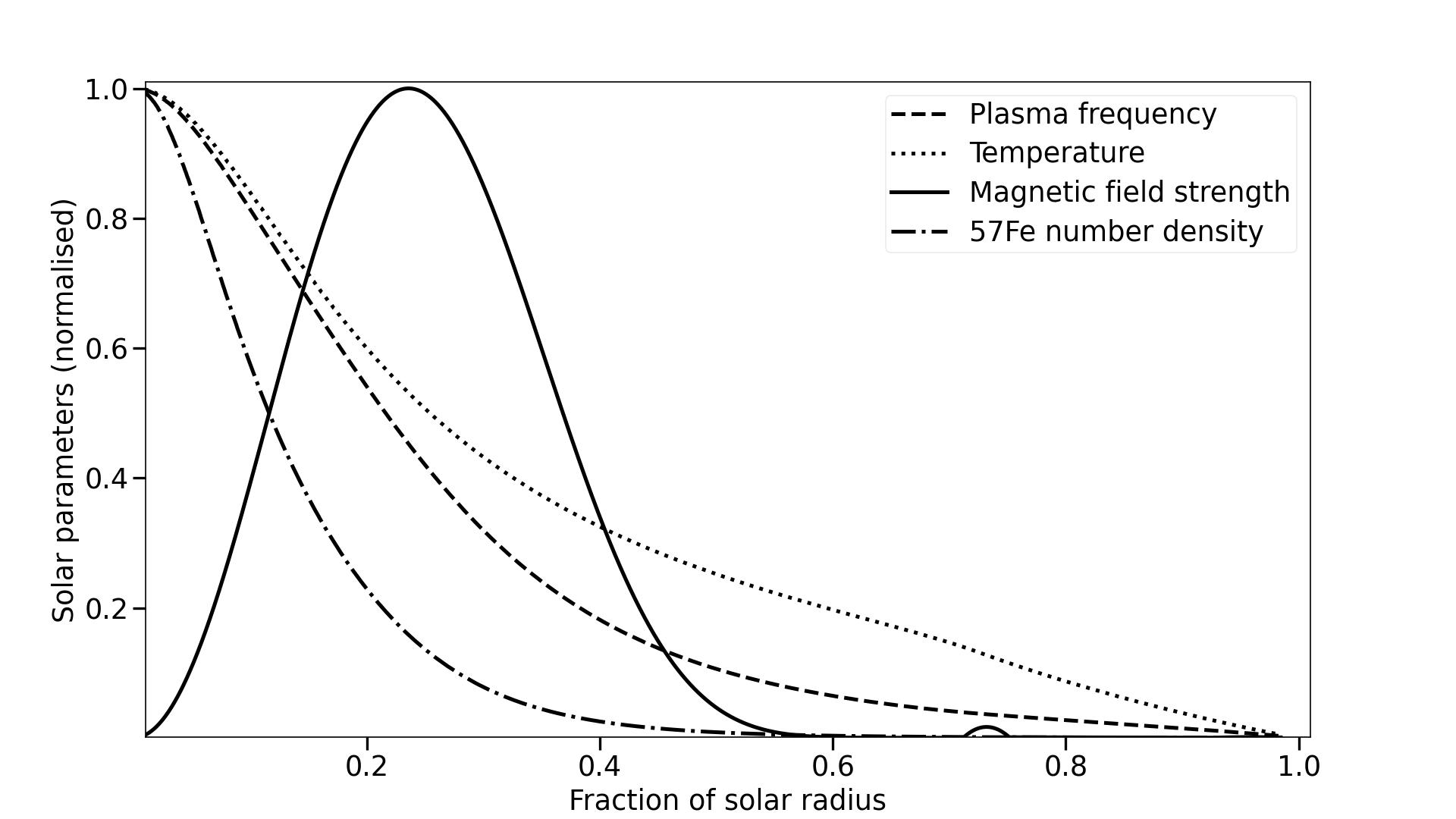}
    \caption{Key solar parameters used in the analysis as a function of the radius, given in units of the solar radius $R_\odot$. The solar parameters are normalised to their maximum values of $\omega_{\operatorname{p}} = 291$ eV, $T = 1335$ eV, $n_{\operatorname{57Fe}} = 2.75\times 10^{24}$ m$^{-3}$, $B = B_0$, where $B_0\in [200,3000]$T~\cite{Bfields} as discussed in the text.}
    \label{fig:solarmodel}
\end{figure}

For the tachocline and outer layer magnetic fields, $a(r)$ takes the form
\begin{equation}
    a(r) = \left( 1 - \left( \frac{r - r_0}{d} \right)^2 \right) B_0
    \label{eq:tachocline}
\end{equation}
for the region $\left| r - r_0 \right| \leq d$.
The values used were $r_0 = 0.732 R_\odot$, $d = 0.02 R_\odot$ for the tachocline magnetic field, and $r_0 = 0.96 R_\odot$, $d = 0.035 R_\odot$ for the outer layers.
The range of values for $B_0$ in these regions was taken to be $B_0\in [4,50]$T for the tachocline, and $B_0\in [3,4]$T for the outer solar layers~\cite{Hoof:2021mld}.
The magnetic field strength as a function of solar radius is shown in Fig. \ref{fig:solarmodel}.

The fluxes of back-converted photons detected by IAXO from this source are shown in the `L' lines for the vacuum and gas run in Fig.~\ref{fig:comparison}. The back-conversion probabilities from Section~\ref{sec:detection} and Section~\ref{sec:gas} and the AGSS09 solar model~\cite{solarModel} were used. Key parameters of the solar model are shown in Fig.~\ref{fig:solarmodel}.
We see that even using the upper limits for the solar magnetic fields, the flux of DPs from this production method is insignificant compared to the thermal production of T-DPs as described in Section \ref{sec:production}.

\section{Dark photon flux from nuclear sources}
\label{sec:pp}

Aside from the thermal processes considered so far, solar DPs can also be produced in nuclear processes. 
Here, we consider DP production in the pp chain and in a number of nuclear de-excitation processes from low energy exited states,
in  a similar  fashion to  what done for BabyIAXO in  Ref.~\cite{DiLuzio:2021qct}.

We begin with the solar proton-deuterium fusion process $p + D \rightarrow  ^3$He$ + \gamma$, which has been extensively studied as a source of axions~\cite{CAST:2009klq,Borexino:2012guz,Bhusal:2020bvx,Lucente:2022esm} as well as of DPs~\cite{DEramo:2023buu}. 
The 5.49 MeV energy produced in the reaction is carried by a photon, which is quickly reabsorbed in the medium. 
However, assuming a 
non-zero kinetic mixing parameter $\chi$, the emitted photon is a superposition of photon-like and sterile-like propagation states as described in Section~\ref{sec:production}. 
The photon-like state is quickly absorbed but the sterile-like state can escape from the sun without thermalising.
As a result, one should expect a nearly monochromatic flux of transversely polarised DPs on Earth (Cf. Ref.~\cite{DEramo:2023buu})
\begin{equation}
    \Phi \approx \frac{\phi}{4 \pi d_\odot^2} \frac{(\omega^2 - m^2)^\frac{3}{2}}{\omega^3} \frac{\chi^2 m^4}{(m^2 - m_{\gamma}^2)^2 + (\omega \Gamma)^2}
    \label{eq:ppFlux}
\end{equation}
where $\phi \approx 1.7 \times 10^{38}$ s$^{-1}$ is the rate of the fusion process and $\omega = 5.49$ MeV.
Here, the photon energy is high enough that $\Gamma$ is dominated by Compton scattering, given by \cite{axion-electronRedondo, Redondo:2008aa}
\begin{equation}
    \Gamma \approx \omega_{\operatorname{p}}^2 \frac{2 \alpha}{3 m_e}.
    \label{eq:compton}
\end{equation}
which is just the $\Gamma_{\operatorname{plas}}$ term in \eqref{eq:pifree}. Note that there is also a flux of L-DPs from this process, but as their production is suppressed compared to that of T-DPs~\cite{DEramo:2023buu} and they are invisible to IAXO they are not considered here.
The spectrum is assumed to be Gaussian with a sharp peak at $\omega = 5.49$ MeV and a width of $\sim 1$ keV due to the thermal motion of the proton. 
Assuming the solar parameters given by AGSS09 for the centre of the sun and using the back-conversion probability given by Eq.~\eqref{eq:Tconversion}, the flux of back-converted photons in IAXO is shown by the line `pp' in Fig.~\ref{fig:comparison}. 
As can be seen, the flux is very low compared to the thermal DP flux, but it is centered in a much higher mass region. 
Thus, a detector sensitive to MeV energies could allow the detection of the DP flux from the pp-chain and 
greatly increase the range of masses accessible to IAXO.\footnote{Notice, however, that this region of the parameter space is already constrained by the XENON experiment (at lower masses) and by cosmological considerations and collider bounds (at higher masses), as shown in Fig.~\ref{fig:oldLimits}.}

Another source of non-thermal solar DP is from the fusion reaction $p + p \rightarrow \, D + e^+ + \nu_e$, which represents the first step of the pp-chain. 
The positron created in this process annihilates very rapidly with an electron in the solar plasma, emitting 2 photons.
In a similar fashion to what discussed before, a non-vanishing photon-DP mixing opens up the possibility that one of the emitted photons converts into a DP and escapes the sun.
The cross section of the $p + p$ process, which is governed by the weak interaction, is far lower than that of the $p + D$ process, which occurs almost immediately once a deuterium nucleus has been produced. 
This means the rate of $p + D \rightarrow ^3$He$ + \gamma$ is limited by the rate of $p + p \rightarrow \, D + e^+ + \nu_e$ and we can assume that $\phi \approx 1.7 \times 10^{38}$ s$^{-1}$ is the same for both. 
This gives a very similar total flux to the above case, albeit with a different spectrum. 
Straightforward kinematics considerations on the $p + p \rightarrow \, D + e^+ + \nu_e$ process, assuming $D$ to be produced at rest, show that the positron emerges with a 96 keV kinetic energy, in the solar reference frame. 
This causes the shift and broadening of the spectrum of emitted photons.
Consequently, the DP is produced with a peak energy of $\omega = 559$ eV and a width of around 100 eV. 
With this in mind, the flux of DPs coming from $e^+ e^-$ annihilation is also given by equation \eqref{eq:ppFlux}, where the only change is that $\omega = 559$~keV.
The total flux is shown in Fig.~\ref{fig:comparison}, with the line labeled `ee'.

Besides nuclear fusion, gamma transitions in nuclear deexcitation can also produce DPs with monochromatic energies. 
The most well known candidate considered in the past, for the axion case, was the M1 nuclear transition of $^{57}$Fe from its first excited state, at 14.4 keV, to the ground state~\cite{Moriyama:1995bz,CAST:2009jdc,DiLuzio:2021qct}.
This transition is particularly intersting for a set of reasons. 
First, $^{57}$Fe is relatively abundant in the sun.
Moreover, the 14.4 keV excitation level is at an energy sufficiently low to be excited in the solar core, where $T\sim$ keV, and it has a short decay time. 
Finally, the fact that this is an M1 transition plays a crucial role for axions which, being pseudoscalar, cannot be produced in electric transitions~\cite{Donnelly:1978ty}.
In the DP case we are not restriced to magnetic transitions.
However, the other criteria still select this transition as the most efficient nuclear deexitation process in the production of a DP flux (Cf. Tab.~\ref{tab:isotopes} in Appendix~\ref{app:isotopes}). 
The probability for a nucleus to be excited to a given energy level by the solar plasma is given by the Boltzmann distribution (Cf. Ref.~\cite{CAST:2009jdc})
\begin{equation}
    p \approx \frac{2 J_1 + 1}{2 J_0 + 1}e^{-\omega / T}
    \label{eq:boltzmann}
\end{equation}
where $J_1$ and $J_0$ are the total angular momentum quantum number for the first excited state and the ground state respectively, $\omega$ is the energy of the transition and $T$ is the temperature of the solar plasma. The rate of photon emission $\phi$ is then given by 
\begin{equation}
    d\phi = \frac{n_I p}{\tau_\gamma} d^3r
    \label{eq:isotopePhotonRate}
\end{equation}
where $n_I$ is the number density of the isotope in question. The lifetime of the exited state to decay by gamma emission is $\tau_\gamma = \tau (1+\alpha_{\operatorname{ic}})$ where $\tau$ is the total lifetime of the state and $\alpha_{\operatorname{ic}}$ is the internal conversion coefficient.
The flux of DPs from this source arriving to Earth is given by
\begin{equation}
    \Phi \approx \frac{\chi^2 m^4}{\tau_{\gamma} d_\odot^2} \frac{2 J_1 + 1}{2 J_0 + 1} \frac{(\omega^2 - m^2)^{3/2}}{\omega^3} \int_0^{R_{\odot}} \frac{n_I r^2 dr}{(m_{\gamma}^2 - m^2)^2 + (\omega \Gamma)^2} e^{-\omega / T}.
    \label{eq:FeFlux}
\end{equation}
The flux of 14.4 keV DPs coming from the $\frac{3}{2}^- \rightarrow \frac{1}{2}^-$ transition of $^{57}$Fe \cite{57FeData} is shown in the curve labeled `Fe' in Fig. \ref{fig:comparison}. Although only the $^{57}$Fe flux is shown, a number of potential isotopes were considered with $^{57}$Fe providing the highest flux. A comparison of various isotopes is shown in Appendix \ref{app:isotopes}.

\section{Contribution from longitudinal dark photons}
\label{sec:Lpure}

Filling the IAXO bores with a gas, for example $^4$He, makes it possible for longitudinal oscillations to be excited by L-DPs, which could lead to their direct detection. 
We remark that both the flux and the conversion probability would be enhanced by a factor $\omega^2 / m^2$ in the case of L-DPs, with respect to the T-DP case. 
Thus, this possibility should be considered in a full assessment of the IAXO potential to detect solar DPs.
In the standard IAXO setup, there is no conversion of L-DPs. This is clear from Eq.~\eqref{eq:dielectic}, where the mixing terms are proportional to $\sin(\theta)\cos(\theta)$, so for the magnetic field orientated perpendicular to the $\mathbf{k}$ direction there is no mixing.
Assuming a different orientation the flux of back-converted photons from L-DPs can be calculated, but this is found to be negligible compared to the flux from T-DPs.

\begin{figure}
    \centering
    \includegraphics[width=\textwidth,keepaspectratio]{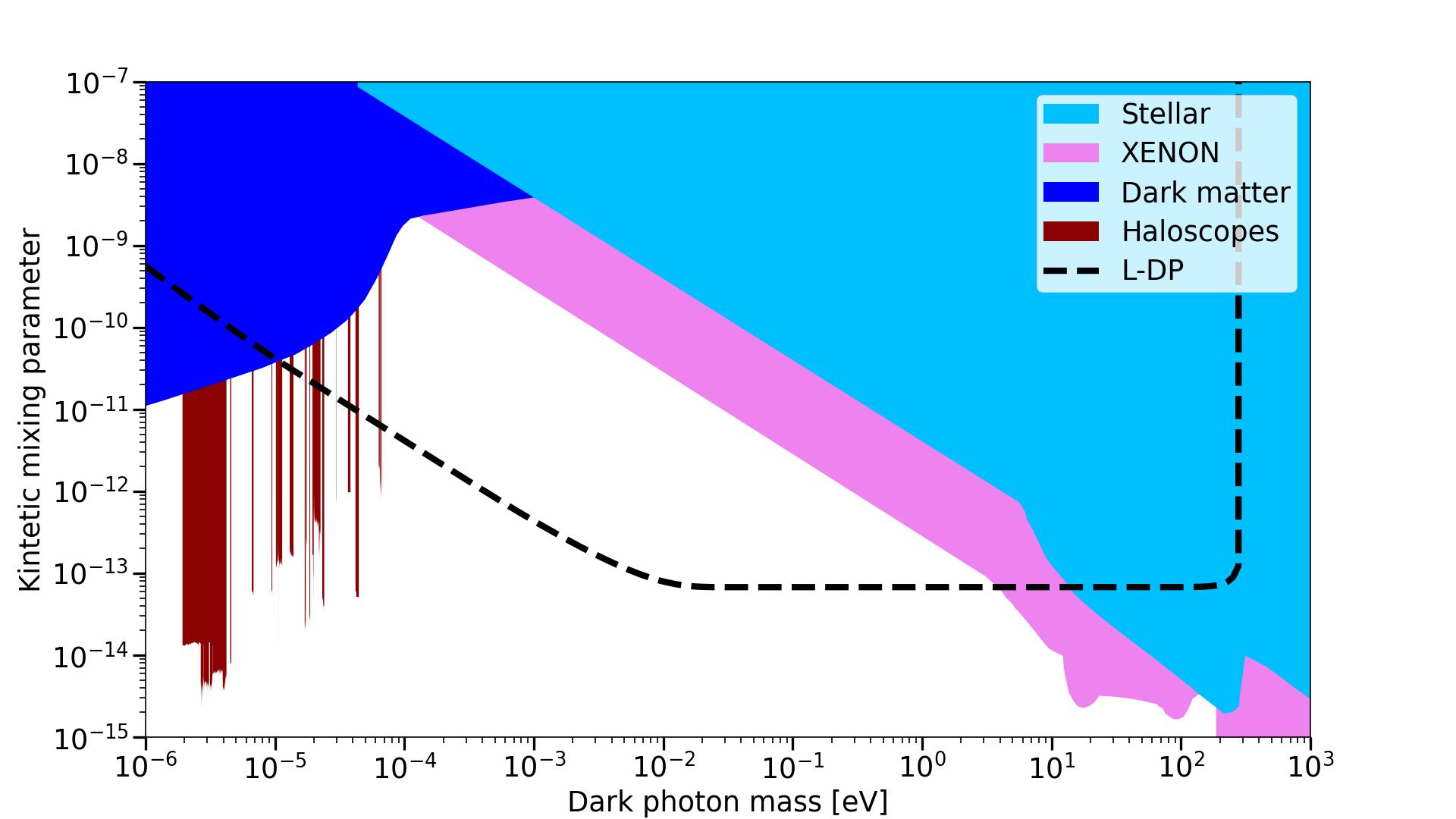}
    \caption{Illustration of the parameter space that would be accessible to an IAXO-like detector modified to be sensitive to longitudinally polarised dark photons (dashed line labelled `L-DP').}
    \label{fig:pureL}
\end{figure}

It is worth considering the more general case of the helioscope detection of L-DPs that are not reliant on mixing through the magnetic field. 
This could be achieved for example with the use of anisotropic crystals rather than a buffer gas, or by a change in the detection method to directly detect the electric field produced by L-DPs. 
The dominance at low $m$ of directly detected L-DPs makes them an interesting potential target for future experiments. 
In Fig.~\ref{fig:pureL}, we show the parameter space that would be accessible to IAXO assuming 
longitudinal oscillations produced resonantly in the detector medium and detected with $100 \%$ efficiency, with conversion probability
\begin{equation}
    P_{S_l \rightarrow A_l} = \frac{\chi^2 m^2}{\Gamma^2} \left( 1 + e^{-\Gamma L} - 2e^{-\Gamma L / 2} \right)\,.
    \label{eq:lConversion}
\end{equation}
The plot assumes 
that the energy range 30 eV - 300 eV has been probed with 5 days on each value of $m_\gamma$, and that all other detector parameters are the same as IAXO. 
Notice that this line should not be considered as indicative of the parameter space that IAXO will be able to probe, but rather a speculative sketch of the region that could be accessed by the helioscope detection of L-DPs, either by a method mentioned above or by some other technique. 
Further considerations which could lead to the exploration of this are left for future works.

\section{Conclusions}
\label{sec:conc}

In this paper we have presented a detailed analysis of the thermal production of DP in the sun, including both the transverse and the longitudinal mode, and considering contributions from both free and bound electrons and ions.  
We have included, for the first time, plasma anisotropic effects generated by the macroscopic magnetic field hosted in the sun. 
The anisotropic plasma allows for conversion of longitudinal DP into transverse modes, which can then be detected with a helioscope set up. 

Non-thermal production channels were also considered, specifically the flux originating from nuclear fusion, $e^+ e^-$ annihilation during the pp-chain, and nuclear de-excitations. 
As the dominant fusion process in the sun, the pp-chain was analysed both as a direct source of X-ray photons and as a positron source. 
Among the nuclear de-excitation processes, $^{57}$Fe was also found to be the dominant solar source of gamma radiation. 
In more massive stars, the CNO cycle dominates over the pp-chain and the different compositions and temperatures may facilitate the excitation of different nuclei, possibly providing additional DP sources. The study of this is left for future works.

When looking at the detection potential, we found that an IAXO run in its basic configuration, using a vacuum in the bores, would improve substantially on the previous CAST bounds, but would not be competitive with the previous limits set by the XENON experiment.
Somewhat surprisingly, the sensitivity for $m \lesssim 0.1$ eV is less in the IAXO gas run than for the vacuum run. The reason is that the photon absorption in the $^4$He gas surpasses the enhancement due to the resonance effect from the effective photon mass in the buffer gas.
The gas run sensitivity would be far greater than that of the vacuum run for $m > 1$ eV. 
However, the required gas pressures are unattainable in IAXO.
It was also found that the detection potential is strongly affected by the lower threshold of detectable energies, and that lowering the threshold from 1 keV to 1 eV would increase the vacuum run sensitivity greatly for dark photon masses less than 1 keV. In terms of increasing the IAXO sensitivity, lowering the energy threshold is more effective than adding a buffer gas for $m  \lesssim 0.1$ eV. However, even with a threshold of 1 eV the IAXO sensitivity falls within the parameter space excluded by the XENON experiment.

In a similar way to the anisotropic effects from the solar plasma, the back-conversion of L-DPs into photons in IAXO was considered assuming the bores are filled with a buffer gas and the magnet is active. 
The photon flux at the detector from this process was found to be zero in the normal IAXO magnetic configuration, and much lower than that from T-DPs even with a modification to allow this conversion. 
However, since the flux of solar L-DPs is much greater than that of T-DPs, for $m \lesssim 1$ eV, a modification to the helioscope design to somehow facilitate the detection of L-DPs would be of great interest. 

To  conclude, we remark that the contents of this paper are currently being implemented in the REST framework~\cite{rest}, which when complete will allow the calculation of a more accurate limit through ray tracing.

\appendix

\section{Solar refraction and absorption}
\label{app:Gamma}
In this appendix we detail the various contributions to the real and imaginary parts of the polarisation tensor in the sun as detailed in Section \ref{sec:production}. We follow very closely the notation of Ref.~\cite{Redondo:2015iea}, to which the reader is referred for more information. In this appendix all symbols retain their definitions from the main paper unless otherwise specified. As briefly mentioned in Section~\ref{sec:production}, at photon energies well above the ionisation energies of the atoms that make up the solar plasma, $\pi_t$ is dominated by free electron effects as all electrons can be treated as free. This is also true of the hottest parts of the solar core, where the atoms are almost completely ionised so only free electrons and ions remain. However we have considered energies down to 1 eV, so bound-free and bound-bound transitions play an important role in the refraction and absorption of photons at low energies. We present here $\pi_t$ decomposed into 4 components
\begin{equation}
    \pi_t = \pi_{\operatorname{plas}} + \pi_{\operatorname{ff}} + \pi_{\operatorname{bf}} + \pi_{\operatorname{bb}}\,,
    \label{eq:piDecomposition}
\end{equation}
where $\pi_{\operatorname{plas}}$ is the contribution from the plasma frequency and Thomson scattering, $\pi_{\operatorname{ff}}$ is the free-free contribution, $\pi_{\operatorname{bf}}$ is the bound-free contribution and $\pi_{\operatorname{bb}}$ is the bound-bound contribution.

The plasma component is given by
\begin{equation}
    \pi_{\operatorname{plas}} = \omega_{\operatorname{p}}^2 \left(1 - i \omega \frac{ 2\alpha}{3 m_e} \right)
    \label{eq:pifree}
\end{equation}
with $\omega_{\operatorname{p}}^2$ defined in Eq.~\eqref{eq:plasmaFreq}. The contribution to the photon mass here is just the plasma frequency, and the contribution to the absorption comes from Thomson scattering.

The free-free contribution $\pi_{\operatorname{ff}}$ comes from electron-ion scattering processes. The real part turns out to be negligible~\cite{Redondo:2015iea}, but the leading contribution to photon absorption near the solar core comes from electron-proton scattering in the process $\gamma + e^- + p^+ \rightarrow e^- + p^+$. This contribution can be written as
\begin{equation}
    \pi_{\operatorname{ff}} = -i\omega \frac{64\pi^2\alpha^3}{3m_e^2\omega^3}\sqrt{\frac{m_e}{2\pi T}} \left(1 - e^{-\omega/T} \right) F \left(\frac{\omega}{T} \right) n_e n_p \,,
\end{equation}
where $n_p$ is the proton number density. $F(\omega/T)$ is the thermally-averaged Gaunt factor to correct the classical results. We take the Born-Elwert approximation~\cite{Elwert:1939} used in Ref.~\cite{Redondo:2015iea}
\begin{equation}
    F(w) = \int_0^\infty \frac{dx}{2} e^{-x^2} \sqrt{x^2 + w} \frac{1 - \exp \left(-2\pi\alpha \sqrt{\frac{m_e}{2T(x^2 + w)}} \right)}{1 - \exp \left(-2\pi\alpha\sqrt{\frac{m_e}{2 T x^2}} \right)} \int_{\sqrt{x^2 + w}-x}^{\sqrt{x^2 + w}-x} \frac{t^3 dt}{(t^2 + y^2)^2}\,.
\end{equation}
We have defined $y \equiv k_D /\sqrt{2 m_e T}$ with Debye screening scale $k_D^2 = 4 \pi \alpha \sum_i Q_i^2 n_i / T$ summing over all charged particles, which we have restricted to electrons and protons.

For atoms that are not fully ionised, the photoelectric effect for outer shell electrons $\gamma + \operatorname{H}^* \rightarrow e^- + p^+$ , or bound-free interaction, has a contribution to both the real and imaginary parts of the polarisation tensor. The contribution to $\Gamma$ can be expressed as
\begin{equation}
    \Gamma_{\operatorname{bf}} = \frac{8\pi m_e \alpha^5}{3\sqrt{3}\omega^3} \left(1-e^{-\omega/T} \right) n_{\operatorname{H}^0} \sum_n \mathcal{Z}_n \frac{1}{n^5} F_{\operatorname{bf}} \Theta(\omega - E_n) + \left(1-e^{-\omega/T} \right) n_{\operatorname{H}^-} \sigma_{\operatorname{H}^-}\,,
    \label{eq:Gammabf}
\end{equation}
and by the Kramers-Kronig relations the real part can be found
\begin{equation}
    m_{\gamma, {\operatorname{bf}}}^2 = \frac{8\pi m_e \alpha^5}{3\sqrt{3}\omega^2} \left(1-e^{-\omega/T} \right) n_{\operatorname{H}^0} \sum_n \mathcal{Z}_n \frac{1}{n^5} \left(\frac{\omega^2}{E_n^2} - \ln \left(\frac{E_n^2}{|E_n^2 - \omega^2|} \right) \right)\,.
    \label{eq:mbf}
\end{equation}
Here $n_{\operatorname{H}^0}$ refers to the number density of neutral hydrogen and $n_{\operatorname{H}^-}$ to that of the negatively charged hydrogen ion. $n$ is the principal quantum number and the ionisation energy is $E_n = \operatorname{Ry}/n^2$. $F_{\operatorname{bf}}$ is the thermally-averaged Gaunt factor for this process which can be ignored since $F_{\operatorname{bf}} \sim 1$ for the interactions of interest here. $\sigma_{\operatorname{H}^-}$ is the cross section for the interaction $\gamma + \operatorname{H}^- \rightarrow \operatorname{H} + e^-$ (Cf. Ref.~\cite{Ohmura:1960})
\begin{equation}
    \sigma_{\operatorname{H}^-} = \frac{\gamma k_e^3}{(1 - \gamma \lambda)(\gamma^2 + k_e^2)^3} \cdot 6.8475\times10^{-18} \operatorname{cm}^2
    \label{eq:sigma}
\end{equation}
where $\gamma = 0.2355883$, $\lambda = 2.646$ and $k_e$ is the wavenumber of the ejected electron in Hartree atomic units (that is, in units of 2 Ry=27.2 eV). 
$\mathcal{Z}_n$ is the probability of finding the atom in state $n$ using the Hummer-Mihalas partition function~\cite{Hummer:1988dva}
\begin{subequations}
\label{eq:HMpartitionfunction}
\begin{equation}
    \mathcal{Z}_n = \frac{2n^2w_ne^{E_n/T}}{\tilde{\mathcal{Z}}}\,,
    \label{eq:Zn}
\end{equation}
\begin{equation}
   \tilde{\mathcal{Z}} = \sum_n2n^2w_ne^{E_n/T}\,.
    \label{eq:Zntilde}
\end{equation}
\end{subequations}
$w_n$ are the occupation probabilities which account for the perturbations of the atomic energy levels due to the electric fields of the plasma ions. Using a Holtsmark distribution for the magnitude of the electric fields, the occupation probabilities can be expressed as
\begin{equation}
    w_n = Q \left(\frac{K_n E_n^2}{4\alpha^2} \left(\frac{3}{4\pi n_p} \right)^{\frac{2}{3}} \right) \,,
    \label{eq:occupationProbs}
\end{equation}
where
\begin{equation}
    K_n = \frac{16 n^2 (n + \frac{7}{6})}{3(n+1)^2(n^2+n+\frac{1}{2})}
    \label{eq:Kn}
\end{equation}
and $Q(\beta)$ is the cumulative distribution function for the Holtsmark distribution, which can be expressed as the sum (Cf. Ref.~\cite{Hummer:1986a})
\begin{subequations}
    
\begin{equation}
    Q(\beta) = \frac{4}{9\pi} \beta^3 \sum_{n=0}^\infty b_n \beta^{2n}\,,
\end{equation}
\begin{equation}
    b_n \equiv (-1)^n \frac{3}{2n+3} \frac{\Gamma \left(\frac{4}{3}n + 2\right)}{\Gamma \left(2n+2 \right)}\,,
\end{equation}
\end{subequations}
where $\Gamma(x)$ is the Gamma function.

Finally, we need to consider bound-bound contributions arising from the excitation and deexcitation of bound electrons. For neutral hydrogen this is given by
\begin{equation}
    \pi_{\operatorname{bb}} = \frac{4\pi\alpha}{m_e}n_{\operatorname{H}^0}\sum_n \mathcal{Z}_n \sum_{n'} f_{n n'} \frac{\omega^2 (\omega^2 - \omega_r^2 - i \omega \gamma_r)}{(\omega^2 - \omega_r^2)^2 + (\omega \gamma_r)^2}\,,
    \label{eq:pibb}
\end{equation}
where $\omega_r = E_n - E_{n'}$ are the resonant frequencies, $\gamma_r = 2\alpha\omega\omega_r/(3m_e)$ are the widths, and $f_{n n'}$ are the oscillator strengths, which can be found at Ref~\cite{NIST_ASD}. Here we only consider the contribution from neutral hydrogen as the most relevant contribution~\cite{Redondo:2015iea}.

From these contributions we can build a model for $m_\gamma^2$ and $\Gamma$ in the sun. For the refractive part there are plasma, bound-free and bound-bound components, and for the absorption there are plasma, free-free, bound-free and bound-bound. As mentioned, we restrict ourselves to hydrogen for the bound-free and bound-bound cases. Overall we find
\begin{subequations}
\label{piFull}
\begin{equation}
    m_\gamma^2 = m_{\operatorname{plas}}^2 + m_{\operatorname{bf}}^2 + m_{\operatorname{bb}}^2\,,
    \label{eq:mGammaFull}
\end{equation}
\begin{equation}
    \Gamma = \Gamma_{\operatorname{plas}} + \Gamma_{\operatorname{ff}} + \Gamma_{\operatorname{bf}} + \Gamma_{\operatorname{bb}}\,.
    \label{eq:GammaFull}
\end{equation}
\end{subequations}

\section{Buffer gas absorption}
\label{app:buffergas}

As for the solar plasma, for the $^4$He buffer gas in the IAXO bores the 2 most important features for this analysis are the effective photon mass in the bores and the absorption of the photons by the gas. 
For the effective mass, we assume that if the photon energy is well above the ionisation energies of $^4$He then the photon interacts with bound electrons and nuclei as if they were free, and the effective photon mass is given by the effective `plasma frequency' of the gas
\begin{equation}
    m_\gamma^2 \approx \omega_{\operatorname{p}}^2 = \frac{4\pi\alpha n_e}{m_e}\,,
    \label{eq:mGammaBufferGas}
\end{equation}
where here $n_e$ is the number density of bound electrons in the gas. 
As the photon energy decreases there comes a point where the refractive index of the gas becomes greater than 1 
and an effective mass can no longer be defined meaning no resonant conversion can occur.

\begin{figure}
    \centering
    \includegraphics[width=\textwidth,keepaspectratio]{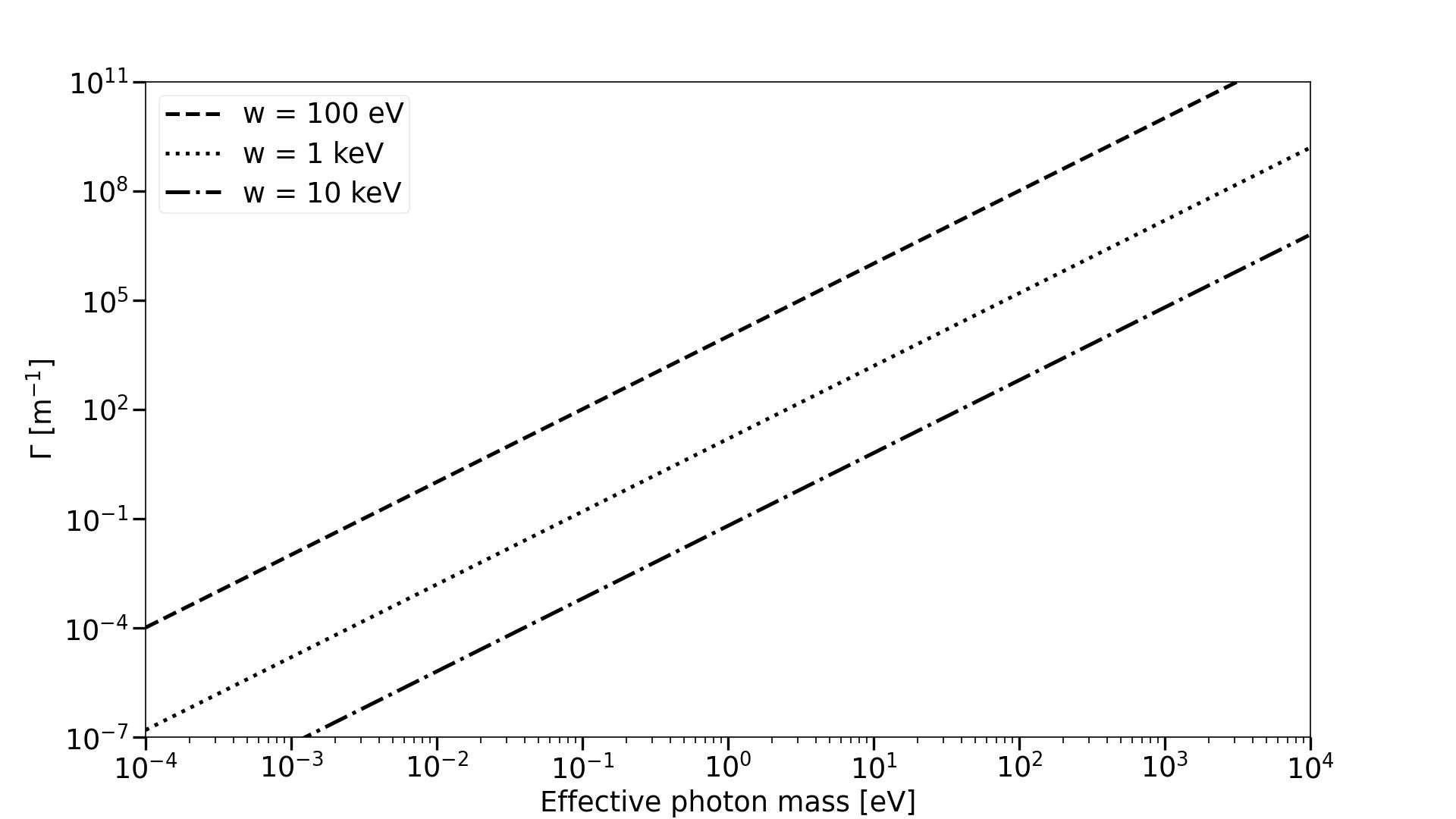}
    \caption{A plot of the IAXO buffer gas absorption length as a function of effective photon mass based on Eq.~\eqref{eq:mGammaBufferGas}.}
    \label{fig:GammaBufferGas}
\end{figure}

The photon absorption length $\Gamma$ as a function of photon energy and gas pressure can easily measured in a lab. Ref.~\cite{juliaThesis} used a fit formula using data from the NIST database~\cite{NIST} to describe the absorption in CAST, and we have adjusted this to define the following for IAXO
\begin{multline}
    \log_{10}\Gamma(\omega,p) = 0.014\log_{10}^6(\omega) + 0.166\log_{10}^5(\omega) + 0.464\log_{10}^4(\omega) + 0.473\log_{10}^3(\omega)\\ - 0.266\log_{10}^2(\omega) - 3.241\log_{10}(\omega) -0.760 + \log_{10}(p) - \log_{10}\frac{T_{\operatorname{IAXO}}}{T_{\operatorname{CAST}}}\,,
    \label{eq:GammaIAXO}
\end{multline}
where photon energy $\omega$ is in keV, gas pressure $p$ is in mbar and absorption length $\Gamma$ is in m$^{-1}$. The CAST temperature $T_{\operatorname{CAST}} =  1.8$ K and the IAXO temperature $T_{\operatorname{IAXO}} = 300$ K are assumed. A plot of the absorption length against the effective photon mass is shown in Fig.~\ref{fig:GammaBufferGas} for relevant photon energies.

\section{Maximum likelihood method}
\label{app:likelihood}

To get the 95\% CL limit on the DP parameter space that can be placed by IAXO assuming no signal events, a maximum likelihood method is used, which we explain in this appendix. We will closely follow the notation of Ref.~\cite{juliaThesis} and all detector parameters are taken from Ref.~\cite{IAXO:2019mpb} unless specified. 

We assume a Poisson distribution for background photons arriving at the detector with a mean value $N_{\operatorname{bg}}$ given by
\begin{equation}
    N_{\operatorname{bg}} = a \epsilon_t t \Phi_{\operatorname{bg}}
    \label{eq:Nbg}
\end{equation}
where $a$ is the sensitive area of the detector, $t$ is the total detection time, $\epsilon_t$ is the detection time efficiency (the proportion of the time $t$ in which IAXO is positioned towards the sun), and $\Phi_{\operatorname{bg}}$ is the background flux. The background flux was assumed to be uniformly distributed over our energy range with a flux equal to the targets set in Ref.~\cite{IAXO:2019mpb}. This was used to simulate the number of events recorded assuming no signal, and a sample of 1000 runs was created and averaged over.

Starting from these premises, the IAXO sensitivity can be estimated using a 
likelihood analysis similar to that found in Ref.~\cite{juliaThesis}. 
Assuming a Poisson process with few events, the log of the likelihood function is given by
\begin{equation}
    \ln(\mathcal{L}) = n ( \ln(\mu) + 1 - \ln(n) ) - \mu
    \label{eq:logL}
\end{equation}
where $n$ is the number of recorded events simulated as described above, and $\mu = N_{\operatorname{bg}} + N_{\operatorname{sig}}(m, \chi)$. 
Here, $N_{\operatorname{sig}}(m, \chi)$ is the expected number of back-converted photons (signal events) as a function of the DP parameters as calculated in the main paper. 
The $95\%$ CL upper limit, $\chi_{\operatorname{lim}}$, can be calculated for each value of $m$ by performing the integral
\begin{equation}
    95\% = \frac{ \int_{0}^{\chi_{\operatorname{lim}}} \mathcal{L}(m, \chi) \,d\chi }{ \int_{0}^{\infty} \mathcal{L}(m, \chi) \,d\chi }
\end{equation}
to give the limit on the parameter space accessible by IAXO, or the $95\%$ CL exclusion of DP parameter space assuming no signal events in the real IAXO run.

\section{Solar isotope table}
\label{app:isotopes}
In this appendix we provide the parameters for the DP flux from the de-excitation of low energy excited nuclear states. 
In all cases, combining all the relevant factors, we find a DP flux considerably smaller than the 14.4 keV flux from the de-excitation of $^{57}$Fe. 

\begin{landscape}
\begin{table}[]
\begin{tabular}{l||lllllllll}
{}             & {$^{57}$Fe}      & {$^{45}$Sc}          & {$^{73}$Ge}            & {$^{83}$Kr}            & {$^{169}$Tm}           & {$^{181}$Ta}           & {$^{187}$Os}           & {{$^{201}$Hg}}         & {$^{235}$U}            \\ \hline \hline
$\omega$ [keV] & 14.41            & 12.40                & 13.28                  & 9.41                   & 8.41                   & 6.24                   & 9.75                   & 1.56                   & 12.98                  \\
$J^\pi_1$      & $\frac{3}{2}^-$  & $\frac{3}{2}^+ $     & $\frac{5}{2}^+$        & $\frac{7}{2}^+ $       & $\frac{3}{2}^+$        & $\frac{9}{2}^-$        & $\frac{3}{2}^-$        & $\frac{1}{2}^-$        & $ \frac{3}{2}^+$       \\
$J^\pi_0$      & $\frac{1}{2}^- $ & $\frac{7}{2}^-$      & $\frac{9}{2}^+$        & $\frac{9}{2}^+$        & $\frac{1}{2}^+$        & $\frac{7}{2}^+$        & $\frac{1}{2}^-$        & $\frac{3}{2}^-$        & $\frac{1}{2}^+$        \\
$t_{1/2}$ [ns]    & 98.3             & $3.26\times10^8$     & {$2910$}               & {156.8}                & {4.09}                 & {$6050$}               & {2.38}                 & {81}                   & {0.5}                  \\
$\alpha_{\operatorname{ic}}$  & 8.56             & 423                  & 497                    & 17.09                  & 263                    & 70.5                   & 280                    & 47000                  & 497                    \\
$A(E)$         & 7.5              & 3.14                 & 3.65                   & 3.25                   & 0.1                    & -0.157                 & 1.4                    & 1.17                  & -0.54                 \\
$\epsilon$     & 0.02119          & 1                    & 0.0773                 & 0.11546                & 1                      & 0.09988                & 0.01271                & 0.1318                 & 0.24286                \\ \hline
$\mathcal{N}$  & 1                & $5.36\times10^{-11}$ & {$7.09\times 10^{-7}$} & {$2.86\times 10^{-3}$} & {$8.94\times 10^{-5}$} & {$5.24\times 10^{-8}$} & {$1.50\times 10^{-5}$} & {$3.77\times 10^{-6}$} & {$1.04\times 10^{-6}$}
\end{tabular}
\caption{The relevant properties of low energy gamma decays of nuclei found in the sun. $t_{1/2}$ is the halflife of the excited state, $A(E)$ is a measure of solar elemental abundance relative to the hydrogen abundance defined as $A(E) \equiv 12 + \log_{10}(n(E)/n(H))$, and $\epsilon$ is the abundance of the particular isotope. $\mathcal{N}$ is a measure of the suppression of the emission rates relative to that of $^{57}$Fe defined as $\mathcal{N} \propto (n_I/t_{1/2}) (1+\alpha)^{-1} e^{-\omega/T}$ and normalised such that $\mathcal{N}(^{57}$Fe$) = 1$. The data on $A(E)$ and $\epsilon$ was taken from Ref.~\cite{asplund2009} and all other data is taken from Ref.~\cite{isotopeData}.}
\label{tab:isotopes}
\end{table}
\end{landscape}

\acknowledgments
This work has been performed as part of the IAXO collaboration. We would like to thank our collaborators, in particular the ones from the IAXO Physics Working Group.
We acknowledge support from the European Union’s Horizon 2020 research and innovation programme under the European Research Council (ERC) grant agreement ERC-2017-AdG788781 (IAXO+). We also acknowledge support from the Spanish Agencia Estatal de Investigación under grant PID2019-108122GB-C31 funded by MCIN/AEI/10.13039/501100011033. We also acknowledge support from the “European Union NextGenerationEU/PRTR” (Planes complementarios, Programa de Astrof\'isica y F\'isica de Altas Energías).
This article/publication is based upon work from COST Action COSMIC WISPers CA21106, supported by COST (European Cooperation in Science and Technology). M.G. and J.R. acknowledge also the Grant PGC2022-126078NB- C21 “Aún más allá de los modelos estándar” funded by MCIN/AEI/ 10.13039/501100011033 and “ERDF A way of making Europe”



\bibliography{report.bib}

\end{document}